\documentclass[12pt,twoside]{article}
\usepackage{amssymb,amsbsy,url}
\newcommand{\be}{\begin{equation}}

\newcommand{\ee}{\end{equation}}
\newcommand{\ba}{\begin{array}}
\newcommand{\ea}{\end{array}}

\let\bi\boldsymbol
\let\ds\displaystyle

\let\ds\displaystyle

\newcommand{\fl}{\hspace*{-7mm}}
\newtheorem{theo}{Theorem}

\begin{document}

\title
{\protect\vspace*{-17mm}\bf\Large The Sasa--Satsuma
(complex mKdV II) and\\ the complex sine-Gordon II equation revisited:\\
recursion operators, nonlocal symmetries, and more}
\author{Artur Sergyeyev\dag\ and Dmitry Demskoi\ddag
\\
\dag\ Mathematical Institute, Silesian University in Opava, \\
Na Rybn\'\i{}\v{c}ku 1,
746\,01 Opava, Czech Republic
\\
\ddag\ School of Mathematics and Statistics,\\ The University of New South
Wales,\\
Sydney, NSW 2052, Australia\\[3mm]
E-mail: {\tt Artur.Sergyeyev@math.slu.cz} and {\tt
demskoi@maths.unsw.edu.au}\vspace{5mm}}
 \maketitle
\vspace*{-15mm}
\begin{abstract}
We present a new symplectic structure and a hereditary recursion operator for
the Sasa--Satsuma 
equation which is widely used in nonlinear optics.
Using
an integro-differential substitution relating this equation
to a third-order 
symmetry flow
of the complex sine-Gordon~II equation
enabled us to find a hereditary recursion operator and higher
Hamiltonian structures for the latter equation. \looseness=-1


We also show that both the Sasa--Satsuma equation
and the third-order 
symmetry flow
for the complex sine-Gordon II equation are bi-Hamiltonian systems,
and we construct several
hierarchies of local and nonlocal symmetries for these systems.

\end{abstract}





\section{Introduction}
Finding a recursion operator for a system of PDEs is of paramount
importance, as the whole integrable hierarchy for the system in
question is then readily generated by the repeated application of
the recursion operator to a suitably chosen seed symmetry, see e.g.\
\cite{bl, olv_eng2, dor} and references therein. Moreover,
using formal adjoint of the recursion operator enables us to
produce infinitely many conserved quantities for our hierarchy (see
e.g.\ \cite{bl, dor} for further details).

The recursion operators of multicomponent systems often
have a richer structure of nonlocalities
than in the one-component case, which makes such
operators more difficult to find.
%
For instance, there are two-component
integrable generalizations, Eqs.(\ref{csg2}) and (\ref{sss}),
of the sine-Gordon equation and of the mKdV equation respectively.
No recursion operator was found for Eqs.(\ref{csg2}) and (\ref{sss}) so far
even though the corresponding Lax pairs \cite{SS, G2,
Sciuto} and bilinear representations \cite{G2, Hir, ghno} are
well known. The goal of the present paper is to fill this gap.
Namely, we find and study the recursion operators for the so-called complex
sine-Gordon II 
equation (\ref{csg2}) and for the Sasa--Satsuma equation
(\ref{sss}).\looseness=-1

The Sasa--Satsuma 
equation \cite{SS} (see also \cite{YO, tp})
for a complex function $U$ has the form
\be\label{cmkdv2}
U_t=U_{xxx}+6 U \bar U U_x+3 U (U \bar U)_x.
\ee
Here and below the bar refers to the complex conjugate.

It is natural to refer to this equation as to the {\em complex mKdV
II} as (\ref{cmkdv2}) is one of the two integrable complexifications
of the famous modified KdV equation, the other complexification
(complex mKdV~I) being simply $U_t=U_{xxx}+6 U \bar U U_x$.

In turn, the complex sine-Gordon II equation \cite{G2, Sciuto} is a
hyperbolic PDE for a complex function $\psi$ of the form
\[
\psi_{xy}=\frac{\bar\psi \psi_x \psi_y}{\psi\bar\psi + c}
+ (2 \psi\bar\psi + c)(\psi\bar\psi + c) k \psi,
\]
where $c$ and $k$ are arbitrary constants.
The usual sine-Gordon equation $\phi_{xy}=c^2 k \sin(\phi)$
is recovered upon setting $\psi=\bar\psi=\sqrt{-c}\sin(\phi/4)$.


The Sasa--Satsuma and the complex sine-Gordon II equation are of
considerable interest for applications. The Sasa--Satsuma equation
is widely used in nonlinear optics, see e.g.\ \cite{kiv, porc} and
references therein, because
the integrable cases of
the so-called higher order nonlinear Schr\"odinger equation \cite{kod}
describing the propagation of short pulses in optical fibers
are related through a gauge transformation either to the Sasa--Satsuma equation \cite{SS}
or to the so-called Hirota equation \cite{Hir73}.
\looseness=-1

The complex sine-Gordon II equation, along with the
Pohlmeyer--Lund--Regge model
\cite{Pohlmeyer,Lund_Regge,G1,Neveu_Papanicolaou}
also known as the complex sine-Gordon I, define integrable perturbations of
conformal field theories \cite{Fateev,Bakas_Park,Brazhnikov,Bakas_Sonnenschein};
see e.g.\ \cite{RS, Sch} for other applications.
Moreover, the complex sine-Gordon~I and II are the only
equations for one complex field in the plane for which
the (multi)vortex solutions are found in closed form \cite{OB,BP,noncoaxial}.

Following \cite{CRC} we set $u=\psi$ and $v=\bar\psi$
and write the complex
sine-Gordon II equation along with its complex conjugate 
as a system for $u$ and $v$:
\be\label{csg2}
\fl u_{xy}=\frac{v u_x u_y}{u
v + c} + (2 u v + c)(u v + c) k u,\qquad v_{xy}=\frac{u v_x v_y}{u v
+ c} + (2 u v + c)(u v + c) k v.
\ee

Likewise, upon setting $p=U$ and $q=\bar U$
in the Sasa--Satsuma 
equation,
proceeding in the same fashion as above and
writing out $(p q)_x$ as $p q_x+p_x q$,
we obtain
\be\label{sss}
p_t=p_{xxx}+9 p q p_x+3 p^2 q_x,\quad
q_t=q_{xxx}+9 p q q_x+3 q^2 p_x.
\ee

From now on we shall 
treat $u$, $v$, $p$ and $q$ as
independent variables that can be real or complex
and consider systems (\ref{csg2}) and (\ref{sss}) that are 
more general than the original complex sine-Gordon~II and Sasa--Satsuma (complex mKdV II)
equations
which can be
recovered under the reductions
$v=\bar u$ and $q=\bar p$ respectively.
In what follows we shall refer to (\ref{csg2})
as to the {\em complex sine-Gordon~II system} and to (\ref{sss}) as to
the {\em Sasa--Satsuma system}.
\looseness=-1


Let us briefly address the relationship of the Sasa--Satsuma system (\ref{sss}) with
other integrable systems.
First of all, Eq.(\ref{sss}) can be obtained \cite{YO} 
as a reduction of the four-component Yajima--Oikawa system (8) from \cite{YO}.

On the other hand, consider the vector modified KdV
equation
\be\label{vmkdv}
V_t=V_{xxx}+\langle V,V\rangle V_x + \langle V,V_x \rangle V
\ee
studied by Svinolupov and Sokolov \cite{svi} (see also \cite{sw, tw}).
Here $V=(V^1,\dots,V^n)^T$ is an $n$-component vector,
$\langle \cdot,\cdot \rangle$
stands for the usual Euclidean scalar product of two vectors,
and the superscript $T$ here and below indicates the transposed matrix.
A simple linear change of variables
$V^1=\sqrt{3}(p+q)/\sqrt{2}$, $V^2=\mathrm{i}\sqrt{3}(p-q)/\sqrt{2}$
takes (\ref{sss}) into (\ref{vmkdv}) with $n=2$.

Setting $r=\sqrt{3}p$ and $s=\sqrt{3}q$ turns (\ref{sss}) into
the system
\begin{equation}\label{fs}
\begin{array}{l}
r_t=r_{xxx}+r (3 s r_x+r s_x),\\[2mm]
s_t=s_{xxx}+s (3 r s_x+s r_x),
\end{array}
\end{equation}
studied by Foursov \cite{f1},
%
who found a Hamiltonian structure for (\ref{fs}) of the form
\be\label{hs} \tilde P=\left( \begin{array}{cc}
-\frac13 r D_x^{-1}\circ r & D_x+\frac13 r D_x^{-1}\circ s\\[2mm]
D_x+\frac13 s D_x^{-1}\circ r & -\frac13 s D_x^{-1}\circ s
\end{array}\right)
\ee where $D_x$ is the operator of total $x$-derivative, see e.g.\
\cite{bl, dor, olv_eng2} for details and for the background on the
recursion operators, Hamiltonian and symplectic structures. The
corresponding Hamiltonian density is \cite{f1} $(2/3)r^2 s^2-r_x
s_x$. This immediately yields a Hamiltonian structure 
(\ref{sshs}) for (\ref{sss}). Note that in \cite{f2,f1} there seems
to be a misprint in $\tilde P$, and in (\ref{hs}) we corrected this
misprint.

In \cite[p.89]{f2} Foursov claims to have found some skew-symmetric
operators that are likely to provide higher Hamiltonian structures
for (\ref{fs}) but he failed to verify that these operators are
indeed Hamiltonian. The explicit form of these operators was not
presented in \cite{f1} or \cite{f2}, so apparently it was never
proved in the literature that (\ref{fs}) (and hence (\ref{sss})) are
bi-Hamiltonian systems.  We establish the bi-Hamiltonian nature of
(\ref{sss}) in Theorem~\ref{roth} below.\looseness=-1

Now 
turn to the complex sine-Gordon II system (\ref{csg2}). Recall that
Eq.(\ref{csg2}) can be obtained (see e.g.\ \cite{G2, CRC,
meshkov92}) as the Euler--Lagrange equation for the functional
$S=\int L \ dx \ dy$, where
\[
L=\frac12 \frac{u_x v_y + u_y v_x}{u v + c} + k (u v + c) u v.
\]
A few conservation laws and generalized symmetries for (\ref{csg2}) can be readily found
e.g.\ using computer algebra \cite{CRC, meshkov92}. In particular \cite{CRC},
Eq.(\ref{csg2}) is compatible with
\be\label{es}
\ba{l}
\ds
u_t=u_{xxx} - \frac{3 u v_x u_{xx}}{u v + c} - \frac{9 u_x^2 v_x}{u v + c}
+ \frac{3 u^2 v_x^2 u_x}{(u v + c)^2},\\[5mm]
\ds v_t=v_{xxx} - \frac{3 v u_x v_{xx}}{u v + c} - \frac{9 v_x^2 u_x}{u v + c}
+ \frac{3 v^2 u_x^2 v_x}{(u v + c)^2}.
\ea
\ee
The compatibility here means that
the flow (\ref{es}) commutes with the nonlocal flow
(\ref{csg2h}) associated with (\ref{csg2}) or, equivalently,
the right-hand sides of (\ref{es}) constitute the characteristic of
a third-order generalized symmetry for (\ref{csg2}),
see e.g.\ \cite{olv_eng2} for general background on symmetries.



The rest of paper is organized as follows. In section II we present
symplectic structure and recursion operator for the Sasa--Satsuma
system (\ref{sss}) and show that (\ref{sss}) is a bi-Hamiltonian
system. In section III we employ a nonlocal change of variables
relating systems (\ref{es}) and (\ref{sss}) in order to construct
recursion operator, Hamiltonian and symplectic structures for
(\ref{csg2}) and (\ref{es}) from those of (\ref{sss}), and we show
that (\ref{es}) is a bi-Hamiltonian system. Finally, in section IV
we discuss the hierarchies of local and nonlocal symmetries for
(\ref{csg2}), (\ref{sss}) and (\ref{es}). \looseness=-1

\section{Recursion operator and symplectic structure\\ for the Sasa--Satsuma system}
%
%
A straightforward but tedious computation proves the following assertion.
\begin{theo}\label{roth} The Sasa--Satsuma system (\ref{sss})
possesses a Hamiltonian structure
\be\label{sshs}
\mathcal{P}=\left(\begin{array}{cc}
-p D_x^{-1}\circ p & D_x+p D_x^{-1}\circ q\\[2mm]
D_x+q D_x^{-1}\circ p & -q D_x^{-1}\circ q
\end{array}\right),
\ee
a 
symplectic structure
\be\label{s4sss}
\fl\mathcal{J}=\left(\begin{array}{cc}
3 p D_x^{-1}\circ p & D_x+5 p D_x^{-1}\circ q\\[2mm]
D_x+5 q D_x^{-1}\circ p & 3 q D_x^{-1}\circ q
\end{array}\right),
\ee
and a hereditary recursion operator
$\mathcal{R}=\mathcal{P}\circ\mathcal{J}$
that can be written as
\be\label{ross}
\ba{l}
\hspace*{-4mm}
\mathcal{R}=\!\left(\!\begin{array}{l@{\hspace{-20mm}}r}
D_x^2+6 p q +q_x D^{-1}\circ p+ b D_x^{-1}\circ p +3 p_x D_x^{-1}\circ b
& 2p^2- 2 z_1 p D_x^{-1}\circ q-3 p_x D_x^{-1}\circ a\\[2mm]
2q^2- 2 z_2 q D_x^{-1}\circ p +3 q_x D_x^{-1}\circ b & D_x^2 +6pq
+p_x D^{-1}\circ q+ a D_x^{-1}\circ q -3 q_x D_x^{-1}\circ a
\end{array}\!\right)\! \ea \ee where $a=p_x+2 z_1 q$, $b=q_x+ 2 z_2
p$; $z_1=D_x^{-1}(p^2)$ and $z_2=D_x^{-1}(q^2)$ 
are potentials for the following conservation laws of
(\ref{sss}):
\[
\fl D_t(p^2)=D_x(2 p p_{xx}-p_x^2+ 6 p^3 q),\quad D_t(q^2)=D_x(2 q q_{xx}-q_x^2+6 p q^2).
\]

Hence (\ref{sss}) has an infinite hierarchy of compatible Hamiltonian structures
$\mathcal{P}_k=\mathcal{R}^k \circ \mathcal{P}$, $k=0,1,2,\dots$,
$\mathcal{P}_0\equiv \mathcal{P}$, an infinite hierarchy of symplectic structures
$\mathcal{J}_k=\mathcal{J}\circ \mathcal{R}^k$, $k=0,1,2,\dots$,
and an infinite hierarchy of commuting
symmetries of the form $\mathcal{K}_i=\mathcal{R}^i(\mathcal{K}_0)$, $i=0,1,2,\dots$, where
$\mathcal{K}_0=(p_x, q_x)^T$.
\looseness=-1


The Sasa--Satsuma system (\ref{sss})
is bi-Hamiltonian with respect to $\mathcal{P}_0$ and $\mathcal{P}_1$: 
\[
\left( \begin{array}{c}
p_t\\[2mm]
q_t\end{array}\right)=\mathcal{P}_0\left( \begin{array}{c}
\delta \mathcal{H}_1/\delta p\\[2mm]
\delta \mathcal{H}_1/\delta q\end{array}\right)=\mathcal{P}_1\left( \begin{array}{c}
\delta \mathcal{H}_0/\delta p\\[2mm]
\delta \mathcal{H}_0/\delta q\end{array}\right),
\]
where $\mathcal{H}_i=\int H_i dx$, $i=0,1$,
$H_0=p q$, $H_1=2 p^2 q^2-p_x q_x$.
\end{theo}
Here $\delta/\delta p$ and $\delta/\delta q$
denote variational derivatives with respect to $p$ and $q$.


Note that the Hamiltonian structure
$\mathcal{P}$ above is nothing but
the second Hamiltonian structure of the AKNS system,
see e.g.\ \cite{magri}.
This Hamiltonian structure (more precisely, its counterpart (\ref{hs}) for Eq.(\ref{fs}))
has already appeared in \cite{f2,f1}.

The symmetries $\mathcal{K}_i$ commute because $\mathcal{R}$ is hereditary.
Applying $\mathcal{R}$ to an obvious symmetry $\mathcal{K}_0=(p_x, q_x)^T$ of (\ref{sss})
yields the symmetry $\mathcal{K}_1=\mathcal{R}(\mathcal{K}_0)=(p_{xxx}+9 p q p_x+3 p^2 q_x,
q_{xxx}+9 p q q_x+3 q^2 p_x)^T$, i.e., the right-hand side of (\ref{sss}). In turn,
$\mathcal{R}(\mathcal{K}_1)$ is a {\em local} fifth-order symmetry for (\ref{sss}).
We guess that $\mathcal{K}_i$
are local for all natural $i$ but so far we were unable to provide a rigorous proof of this.
\looseness=-1


Unlike the overwhelming majority
of the hitherto known recursion operators,
see e.g.\ discussion in \cite{serg2} and references therein, the nonlocal variables
appear explicitly in the coefficients of $\mathcal{R}$.
Perhaps this is the very reason
why $\mathcal{R}$ was not found earlier.
%
The nonlocal variables in the coefficients of
$\mathcal{R}$ are {\em abelian} pseudopotentials
as in \cite{kk} and unlike e.g.\ the nonlocalities in the 
recursion operator 
discovered by Karasu et al.\ \cite{kar}
and later rewritten in \cite{as_sro}:
the nonlocalities in the operator from \cite{kar, as_sro}
are {\em nonabelian} pseudopotentials.

\section{Recursion operator\\ for the complex sine-Gordon~II system\looseness=-1}
%


There is a 
well-known (see e.g.\ \cite{agf} for discussion and references)
transformation $z=\sqrt{2/3}g_x$  relating the symmetry flow
$g_t=g_{xxx}+g_x^3/2$
of the sine-Gordon equation $g_{xy}=\sin(g)$
and the mKdV equation
$z_t=z_{xxx}+z^2 z_x$.
%
Moreover, there exists \cite{agf} a nonlocal generalization
of this transformation that sends
the third-order symmetry flow
\[
\ba{l} u_t=\ds u_{xxx} - 3 \frac{u v_x u_{xx}}{u v + c} + 3
\frac{(-u v
u_x - c u_x + u^2 v_x) u_x v_x}{(u v + c)^2},\\[5mm]
v_t=\ds v_{xxx} - 3 \frac{v u_x v_{xx}}{u v + c} - 3 \frac{(-v^2 u_x
+ u v v_x + c v_x) u_x v_x}{(u v + c)^2} \ea
\]
of the complex sine-Gordon I equation (see e.g.\ \cite{G1,CRC, meshkov92})
\be\label{csg1}
\fl u_{xy}=\frac{v u_x u_y}{u
v + c} + (u v + c) k u,\qquad v_{xy}=\frac{u v_x v_y}{u v
+ c} + (u v + c) k v,
\ee
into the two-component generalization of the mKdV equation
\[
\ba{l} p_t=p_{xxx}+6 p q p_x,\qquad
q_t=q_{xxx}+6 p q q_x \ea
\]
that belongs to the hierarchy of the well-known AKNS system
\be\label{akns}
\ba{l}
p_t=p_{xx}+p^2 q,\qquad 
q_t=-q_{xx}-q^2 p.
\ea
\ee

Note that this nonlocal transformation also sends \cite{agf}
the second-order symmetry flow
\[
\ba{l} u_t=\ds u_{xx} - 2 \frac{u u_x v_x}{u v + c},\qquad 
v_t=\ds -v_{xx} + 2 \frac{v u_x v_x}{u v + c} \ea
\]
of (\ref{csg1})
into the AKNS system (\ref{akns}).

It turns out that upon a suitable redefinition of nonlocal variables
the nonlocal transformation in question in combination with
a suitable rescaling of dependent variables $p$ and $q$
also sends the third-order symmetry flow (\ref{es})
of the complex sine-Gordon II equation (\ref{csg2}) into the Sasa--Satsuma system (\ref{sss}).
Namely, we have the following result.
\looseness=-1
\begin{theo}\label{es2sss}The substitution
\[
p=\frac{\mathrm{i}
\sqrt{2} u_x\exp({-\frac{1}{2} w_1})}{\sqrt{u v + c}},\quad
q=\frac{\mathrm{i}\sqrt{2} v_x\exp({\frac{1}{2} w_1})}{\sqrt{u v + c}},
\]
where $\mathrm{i}=\sqrt{-1}$, $w_1=D_x^{-1}(\rho_1)$ is a potential for
the conservation law $D_t(\rho_1)=D_x(\sigma_1)$ of (\ref{es}),
\[
\ba{l}
\ds
\rho_1=\frac{u v_x-v u_x}{u v + c}, \quad
\sigma_1=\frac{u v_{xxx}-v u_{xxx}}{u v+c}+\frac{(3 u v+ 2 c)(v_x u_{xx}-u_x v_{xx})}{(u v+c)^2}\\[5mm]
\ds+\frac{(12 uv +11c)(v u_x - u v_x)u_x v_x}{(u v+c)^3},
\ea
\]
takes Eq.(\ref{es}) into
the Sasa--Satsuma system (\ref{sss}).
\looseness=-1
\end{theo}\nopagebreak
Here and below $D_x$ and $D_t$ stand for the total $x$- and $t$-derivatives
as defined e.g.\ in \cite{olv_eng2}.

\noindent{\bf Remark 1} {\em Let \[
\tilde p=\frac{\mathrm{i}
\sqrt{2} u_x\exp({-\frac{1}{2} w_1})}{\sqrt{u v + c}},\quad
\tilde q=\frac{\mathrm{i}\sqrt{2} v_x\exp({\frac{1}{2} w_1})}{\sqrt{u v + c}},
\]
where $w_1=D_x^{-1}(\rho_1)$ is now a potential for the conservation
law $D_y(\rho_1)=D_x(\theta_1)$ of (\ref{csg2}), $\rho_1$ is given
above and
\[
\theta_1=\frac{v u_y-u v_y}{u v + c}.
\]
Then for $k=0$ we have
\[
D_y(\tilde p)=0,\qquad D_y(\tilde q)=0,
\]
where $D_y$ stands for the total $y$-derivative.
In other words, if $k=0$ then 
$\tilde p$ and $\tilde q$
provide {\em nonlocal} $y$-integrals for (\ref{csg2}),
and $D_x\ln(\tilde p)$ and $D_x\ln(\tilde q)$ are {\em local} $y$-integrals of (\ref{csg2}).
Therefore, the complex sine-Gordon II system (\ref{csg2}) for $k=0$
is 
Liouvillean and $C$-integrable, see \cite{DS, dem2}
for the construction of symmetries of such systems using the integrals thereof.
Using the above local integrals enables us to find
the general solution for (\ref{csg2}) with $k=0$ along the lines of \cite{zs}.}


%

Passing from $p$ and $q$ to $u$ and $v$ yields from
$\mathcal{R}$ a recursion operator $\mathfrak{R}_0$ for (\ref{es}).
The operator $\mathfrak{R}_0$ is hereditary
because so is $\mathcal{R}$, see \cite{ff}.
It is easily seen that 
$\mathfrak{R}_0$ is a recursion operator for (\ref{csg2}) as well.
Upon removing an inessential overall
constant factor in $\mathfrak{R}_0$ we obtain the following result:
\begin{theo}\label{rothsg} The complex sine-Gordon II system (\ref{csg2}) has
a hereditary recursion operator
\[
\ba{l}
\mathfrak{R}=\left( \begin{array}{c@{\hspace{-8mm}}c}
D_x^2-\ds\frac{2 u v_x}{u v + c}+\ds\frac{u v_{xx}-12 u_x v_x}{u v + c}+\frac{c u_x v_x}{(u v + c)^2}&
\ds\frac{-2 u u_{xx}+6 u_x^2}{u v + c} + \frac{4 u^2 u_x v_x}{(u v + c)^2}\\[5mm]
\ds\frac{-2 v v_{xx}+6 v_x^2}{u v + c} + \frac{4 v^2 u_x v_x}{(u v + c)^2}&
D_x^2-\ds\frac{2 v u_x}{u v + c}+\ds\frac{v u_{xx}- 12 u_x v_x}{u v + c}+\frac{c u_x v_x}{(u v + c)^2}
\end{array}\right)\\[12mm]
+\left( \begin{array}{cc}
\sum\limits_{\alpha=1}^5 Q_\alpha^1 D_x^{-1}\circ {\gamma}_{\alpha,1} &
\sum\limits_{\alpha=1}^5 Q_\alpha^1 D_x^{-1}\circ {\gamma}_{\alpha,2}\\[5mm]
\sum\limits_{\alpha=1}^5 Q_\alpha^2 D_x^{-1}\circ {\gamma}_{\alpha,1} &
\sum\limits_{\alpha=1}^5 Q_\alpha^2 D_x^{-1}\circ {\gamma}_{\alpha,2}
\end{array}\right),
\ea
\]
where $w_1$ is as in Remark 1;
$y_1=D_x^{-1}\left((u_x^2 \exp(-w_1))/(u v + c)\right)$
and $y_2=D_x^{-1}\left((v_x^2 \exp(w_1))/(u v + c)\right)$
are potentials for the nonlocal conservation laws $D_y(\zeta_i)=D_x(\xi_i)$, $i=1,2$, of (\ref{csg2}),
\[
\ba{l}
\zeta_1={\ds\frac{u_x^2 \exp(-w_1)}{u v + c}},\quad \xi_1=k u^2 (u v+c)\exp(-w_1),\\[5mm]
\zeta_2={\ds\frac{v_x^2 \exp(w_1)}{u v + c}},\quad \xi_2=k v^2 (u v+c)\exp(w_1);
\ea
\]
\[
\ba{l}
\boldsymbol{Q}_1=\left( \begin{array}{c}
y_1 u\\[2mm]
u_x\exp(-w_1)-v y_1
\end{array}\right),\quad
\boldsymbol{Q}_2=\left( \begin{array}{c}
v_x\exp(w_1)-u y_2\\[2mm]
y_2 v
\end{array}\right),\\[10mm]
\boldsymbol{Q}_3=\left( \begin{array}{c}
u\\[2mm]
-v
\end{array}\right),\quad
\boldsymbol{Q}_4=c\left( \begin{array}{c}
-u_{xx} + \ds\frac{2 u u_x v_x}{u v + c} +4 \exp(w_1) y_1 v_x- 4 u y_2 y_1\\[10mm]
v_{xx} - \ds\frac{2 v u_x v_x}{u v + c} -4 \exp(w_1) y_2 u_x+ 4 v y_2 y_1\end{array}\right),\quad
\boldsymbol{Q}_5=\left( \begin{array}{c}
u_x\\[2mm]
v_x
\end{array}\right)
\ea
\]
are symmetries for (\ref{csg2}),
and
\[
\ba{l}
\fl\boldsymbol{\gamma}_1=\left({\ds \frac{\exp(w_1) v v_x^2 +c y_2 v_x}{(u v + c)^2},
\frac{\exp(w_1) v_{xx}}{u v + c} -
\frac{2\exp(w_1) v u_x v_x+c y_2 u_x}{(u v + c)^2}}\right),\\[7mm]
\fl\boldsymbol{\gamma}_2=\left({\ds \frac{\exp(-w_1) u_{xx}}{u v + c} -
\frac{2\exp(-w_1) u u_x v_x+c y_1 v_x}{(u v + c)^2}},
{\ds\frac{\exp(-w_1) u u_x^2 +c y_1 u_x}{(u v + c)^2}
}\right)
\ea
\]
\[
\ba{l}
\fl\boldsymbol{\gamma}_3=\left({\ds -\frac{v_{xxx}}{u v + c}
+ \frac{2(v v_x + 2 \exp(-w_1) u v y_2 + 2 \exp(- w_1) c y_2) u_{xx}+(2 v u_x + u v_x) v_{xx}}{(u v + c)^2}
}\right.\\[7mm]
\fl\left.{\ds+\frac{2 v_x u_x(-v^2 u_x + 2 c v_x)}{(u v + c)^3}-\frac{2 v_x(-u_x v_x + 4 \exp(- w_1) u y_2 u_x
+ 2 \exp(w_1) v y_1 v_x + 2 c y_2 y_1)}{(u v + c)^2}},\right.\\[7mm]
\fl\left.{\ds \frac{u_{xxx}}{u v + c}
- \frac{2(u u_x + 2 \exp(w_1) u v y_1 + 2 \exp(w_1) c y_1) v_{xx}-(2 u v_x + v u_x) u_{xx}}{(u v + c)^2}
}\right.\\[7mm]
\fl\left.{\ds-\frac{2 v_x u_x(-u^2 v_x + 2 c u_x)}{(u v + c)^3}+\frac{2 u_x(-u_x v_x + 4 \exp(w_1) v y_1 v_x
+ 2 \exp(-w_1) u y_2 u_x + 2 c y_2 y_1)}{(u v + c)^2}}\right),\\[7mm]
\fl\boldsymbol{\gamma}_4=\left({\ds\frac{v_x}{(u v + c)^2}, -\frac{u_x}{(u v + c)^2}}\right),\quad
\boldsymbol{\gamma}_5=\left({\ds\frac{v_{xx}}{(u v + c)}-\frac{u v_x}{(u v + c)^2}},
{\ds\frac{u_{xx}}{(u v + c)}-\frac{v u_x}{(u v + c)^2}}\right)
\ea
\]
are cosymmetries for (\ref{csg2}).
\end{theo}
Here $Q_\alpha^i$ and $\gamma_{\alpha,j}$
denote $i^{\mathrm{th}}$ component of $\boldsymbol{Q}_\alpha$ and $j^{\mathrm{th}}$
component of $\boldsymbol{\gamma}_\alpha$ respectively, i.e.,
\[
\boldsymbol{Q}_\alpha\equiv \left(\ba{c}Q_{\alpha}^1\\[2mm] Q_{\alpha}^2\ea\right),\quad
\boldsymbol{\gamma}_\alpha\equiv \left(\gamma_{\alpha,1}, \gamma_{\alpha,2}\right).
\]
Note that if we use the tensorial notation (see e.g.\ \cite{meshkov92}),
we can rewrite the operator
\[
\left( \begin{array}{cc}
\sum\limits_{\alpha=1}^5 Q_\alpha^1 D_x^{-1}\circ {\gamma}_{\alpha,1} &
\sum\limits_{\alpha=1}^5 Q_\alpha^1 D_x^{-1}\circ {\gamma}_{\alpha,2}\\[5mm]
\sum\limits_{\alpha=1}^5 Q_\alpha^2 D_x^{-1}\circ {\gamma}_{\alpha,1} &
\sum\limits_{\alpha=1}^5 Q_\alpha^2 D_x^{-1}\circ {\gamma}_{\alpha,2}
\end{array}\right)
\]
in a more concise form, namely
$\sum\limits_{\alpha=1}^5\boldsymbol{Q}_\alpha \otimes D_x^{-1}\circ\boldsymbol{\gamma}_\alpha$.


\noindent{\bf Remark 2}
{\em Eq.(\ref{es}) 
has a recursion operator of precisely the same form as the
$\mathfrak{R}$ given above,
but in this case the nonlocal variables should be defined in a
slightly different way: $w_1$ should be as in Theorem~\ref{es2sss},
and $y_1=D_x^{-1}\left((u_x^2 \exp(-w_1))/(u v + c)\right)$,
$y_2=D_x^{-1}\left((v_x^2 \exp(w_1))/(u v + c)\right)$ should now be
potentials for the nonlocal conservation laws
$D_t(\zeta_i)=D_x(\chi_i)$, $i=1,2$, of (\ref{es}), where $\zeta_i$
are as in Theorem~\ref{roth} and
\[
\ba{l}
\chi_1=\ds \exp(- w_1)\left(\frac{2 u_x u_{xxx}}{u v + c}
- \frac{u_{xx}^2}{u v + c} - \frac{2 u u_x v_x u_{xx}}{(u v + c)^2}
 - \frac{2 u u_x^2 v_{xx}}{(u v + c)^2}
- \frac{2(6 u v + 7 c) u_x^3 v_x}{(u v + c)^3}
+ \frac{3 u^2 u_x^2 v_x^2}{(u v + c)^3}\right),\\[5mm]
\chi_2=\ds \exp(w_1)\left(\frac{2 v_x v_{xxx}}{u v + c}
 - \frac{v_{xx}^2}{u v + c}
 - \frac{2 v v_x^2 u_{xx}}{(u v + c)^2}
 - \frac{2 v u_x v_x v_{xx}}{(u v + c)^2}
+ \frac{3 v^2 u_x^2 v_x^2}{(u v + c)^3}
- \frac{2 (6 u v + 7 c) u_x v_x^3}{(u v + c)^3}\right).
\ea
\]
}

Recall that Eq.(\ref{csg2}) possesses \cite{meshkov92, mo} a local
symplectic structure 
\[
\mathfrak{J}=\left(
\begin{array}{c@{\hspace{-10mm}}c}
0&{\ds\frac{1}{uv+c}} D_x-{\ds\frac{u v_x}{(uv +c)^2}}\\[4mm]
{\ds\frac{1}{uv+c}} D_x-{\ds\frac{v u_x}{(uv +c)^2}}& 0
\end{array}\right),
\]
i.e., Eq.(\ref{csg2}) can be written in the form $\mathfrak{J}\boldsymbol{u}_y=k (2 u v+c)(v, u)^T
=\delta\mathcal{H}/\delta\boldsymbol{u}$, where $\boldsymbol{u}=(u,v)^T$
and $\mathcal{H}=\int k (u v + c) u v dx$.
The symplectic structure $\mathfrak{J}$
is readily seen to be shared by 
Eq.(\ref{es}).

Inverting $\mathfrak{J}$ yields a nonlocal Hamiltonian structure for (\ref{es}) of the form
\[
\fl\mathfrak{P}=\left( \begin{array}{c@{\hspace{-35mm}}c}
0&\exp(-w_1/2) \sqrt{uv+c}\circ D_x^{-1}\circ \exp(w_1/2)\sqrt{uv +c}\\[2mm]
\exp(w_1/2) \sqrt{uv+c}\circ D_x^{-1}\circ \exp(-w_1/2)\sqrt{uv +c}& 0
\end{array}\right),
\]
where $w_1$ should be interpreted as in Theorem \ref{es2sss}. It can
be shown that if we go from the $(u,v)$- to the $(p,q)$-variables
using the transformation from Theorem \ref{es2sss} then we obtain from $\mathfrak{P}$
the Hamiltonian structure 
(\ref{sshs}) for the Sasa--Satsuma system
up to an inessential overall constant factor.


The higher Hamiltonian (resp.\ symplectic) structures for (\ref{es})
are given by the formulas $\mathfrak{P}_n=\mathfrak{R}^n
\circ\mathfrak{P}$ (resp.\ $\mathfrak{J}_n=\mathfrak{J}\circ
\mathfrak{R}^n$), where $n=1,2,3,\dots$. The operator $\mathfrak{R}$ is hereditary,
and hence all of these structures are compatible. However, 
the structures
$\mathfrak{P}_1$ and $\mathfrak{J}_1$
are already very cumbersome, so we do not display
them here. 
Nevertheless, it is readily checked that the following assertion
holds.
\begin{theo}
Eq.(\ref{es}) is a bi-Hamiltonian system:
\[
\left( \begin{array}{c}
u_t\\[2mm]
v_t\end{array}\right)=\mathfrak{P}_0\left( \begin{array}{c}
\delta \mathfrak{H}_1/\delta u\\[2mm]
\delta \mathfrak{H}_1/\delta v\end{array}\right)=\mathfrak{P}_1\left( \begin{array}{c}
\delta \mathfrak{H}_0/\delta u\\[2mm]
\delta \mathfrak{H}_0/\delta v\end{array}\right),
\]
where $\mathfrak{H}_i=\int \tilde H_i dx$, $i=0,1$,
\[
\fl\tilde H_0=-\ds\frac{u_x v_x}{u v+c},\quad
\tilde H_1=\ds \frac{u_{xx} v_{xx}}{u v+c}-\ds\frac{v u_x v_x u_{xx}
+u u_x v_x v_{xx}-4 u_x^2 v_x^2}{(u
v+c)^2}+\frac{u v u_x^2 v_x^2}{(u v+c)^3}.
\]
\end{theo}

It 
is straightforward to verify that
the action of $\mathfrak{R}$
on the 
obvious symmetry $\boldsymbol{u}_x$ of (\ref{es}) yields the
right-hand side of (\ref{es}), and
$\mathfrak{R}^2(\boldsymbol{u}_x)$ is a fifth-order {\em local}
generalized symmetry for (\ref{csg2}). We guess that the repeated
application of $\mathfrak{R}$ to $\boldsymbol{u}_x$ yields a
hierarchy of {\em local} generalized symmetries for~(\ref{csg2}) but
we were not able to prove this in full generality so far, as
presence of the nonlocalities $y_i$ in the coefficients of
$\mathfrak{R}$ appears to render useless the hitherto known ways of
proving locality for hierarchies of symmetries generated by the
recursion operator, cf.\ e.g.\ \cite{serg} and references therein.
Nevertheless, as $\mathfrak{R}$ is a recursion operator for
(\ref{es}), Proposition~2 from \cite{serg2} tells us that the only
nonlocalities that could possibly appear in the hierarchy of the
symmetries $\mathfrak{R}^k(\boldsymbol{u}_x)$, $k=1,2,\dots$, are
potentials of (possibly nonlocal) conservation laws for
(\ref{csg2}). \looseness=-1

\section{Nonlocal symmetries}
First of all, we can consider Eq.(\ref{csg2}) as a nonlocal symmetry
flow of Eq.(\ref{es}). We have already noticed above that Eq.(\ref{csg2})
can be written as
\be\label{csg2s}
\mathfrak{J}\boldsymbol{u}_y
=\delta\mathfrak{H}/\delta\boldsymbol{u},
\ee
where
$\mathfrak{H}=\int k (u v + c) u v dx$.
Acting by $\mathfrak{P}=\mathfrak{J}^{-1}$ on both sides of (\ref{csg2s})
we can formally rewrite the complex sine-Gordon II system (\ref{csg2}) in the evolutionary form $\boldsymbol{u}_y
=\mathfrak{P}(\delta\mathfrak{H}/\delta\boldsymbol{u})$, that is,
\be\label{csg2h}
u_y=k \exp(-w_1/2)\sqrt{u v + c}\ \omega_2,\quad
v_y=k \exp(w_1/2)\sqrt{u v + c}\ \omega_1,
\ee
where $\omega_1=D_x^{-1}\left(v(2 u v + c) \exp(-w_1/2)\sqrt{u v + c}\right)$
and $\omega_2=D_x^{-1}\left(u(2 u v + c) \exp(w_1/2)\sqrt{u v + c}\right)$
are potentials for the following nonlocal conservation laws of (\ref{es}):
\[
\ba{l}
\!\! D_t\left(v(2 u v + c) \exp(-w_1/2)\sqrt{u v + c}\right)
=D_x\left(\exp(-w_1/2)\sqrt{u v + c}
\left(\ds\frac{(4 u v + 3 c) v^2 u_{xx}}{u v + c}
+ (4 u v + c) v_{xx}\right.\right.\\[7mm]
\left.\left.- \ds \frac{2 v^3 u_x^2}{u v + c}
- \frac{v(20 u^2 v^2 + 26 c u v + 7 c^2) u_x v_x}{(u v + c)^2}
- 2 u v_x^2\right)\right),
\ea
\]
\[
\ba{l}
\!\! D_t\left(u(2 u v + c) \exp(w_1/2)\sqrt{u v + c}\right)
=D_x\left(\exp(w_1/2)\sqrt{u v + c}
\left(
(4 u v + c) u_{xx}
+ \ds\frac{u^2(4 u v + 3 c) v_{xx}}{u v + c}\right.\right.\\[7mm]
\left.\left. - 2 v u_x^2
- \ds\frac{u(20 u^2 v^2 + 26 c u v + 7 c^2) u_x v_x}{(u v + c)^2}
- \frac{2 u^3 v_x^2}{u v + c}
\right)\right).
\ea
\]
Eq.(\ref{csg2h}) is compatible with Eq.(\ref{es}), i.e.,
the right-hand side of (\ref{csg2h}) divided by $k$, \[
\boldsymbol{Q}_{-1}=\left(\ba{l}
\exp(-w_1/2)\sqrt{u v + c}\ \omega_2\\[4mm]
\exp(w_1/2)\sqrt{u v + c}\ \omega_1\ea\right),
\]
is a nonlocal symmetry for (\ref{es}). Moreover, we have $\mathfrak{R}(\boldsymbol{Q}_{-1})=c^2 \boldsymbol{u}_x$, i.e., the
action of $\mathfrak{R}$ on $\boldsymbol{Q}_{-1}$ gives the `zeroth' (obvious) symmetry $\boldsymbol{u}_x$ up to a constant factor.
Thus, Eq.(\ref{csg2h}) (and hence the complex sine-Gordon II system (\ref{csg2}))
can be considered as a first negative flow in
the hierarchy of (\ref{es}). Note that if we consider $\mathfrak{R}$ as a recursion operator for Eq.(\ref{csg2}),
we have $\mathfrak{R}(\boldsymbol{u}_y)=k c^2 \boldsymbol{u}_x$, in perfect agreement with the above result.

Eqs.(\ref{sss}) and (\ref{csg2}) possess nonlocal symmetries
of the form
\[
\mathcal{G}_1=\left( \begin{array}{c}
q_x+z_2 p\\[2mm]
- z_2 q\end{array}\right),\quad
\mathcal{G}_2=\left( \begin{array}{c}
-z_1 p\\[2mm]
p_x+z_1 q\end{array}\right),
\]
and
\[
\boldsymbol{Q}_1=\left( \begin{array}{c}
y_1 u\\[2mm]
u_x\exp(-w_1)-v y_1
\end{array}\right),\quad
\boldsymbol{Q}_2=\left( \begin{array}{c}
v_x\exp(w_1)-u y_2\\[2mm]
y_2 v
\end{array}\right),
\]
respectively, where $z_1$ and $z_2$ were defined in Theorem \ref{roth}.
Therefore we have 
four hierarchies of nonlocal symmetries:
$\mathcal{R}^i(\mathcal{G}_j)$ for (\ref{sss})
and 
$\mathfrak{R}^i(\boldsymbol{Q}_j)$
for (\ref{csg2}), where in both cases $j=1,2$ and $i=0,1,2,\dots$.
\looseness=-1

Two more hierarchies of nonlocal symmetries have the form
$\mathcal{R}^i(\mathcal{G}_0)$  for (\ref{sss})
and $\mathfrak{R}^i(\boldsymbol{Q}_0)$
for (\ref{csg2}), where $i=1,2,\dots$,
\[
\mathcal{G}_0=\left( \begin{array}{c}
p\\[2mm]
- q\end{array}\right),\quad 
\boldsymbol{Q}_0=\left( \begin{array}{c}
u\\[2mm]
-v
\end{array}\right).
\]
Two somewhat more `usual' nonlocal hierarchies of master symmetries
(the latter are, roughly speaking, time-dependent symmetries such
that repeatedly commuting them with a suitable
time-independent symmetry
yields an infinite hierarchy of {\em time-independent} commuting
symmetries for the system in question, see
e.g.\ \cite{bl} and references therein for further details),
$\mathcal{R}^i(\mathcal{S}_0)$ for (\ref{sss})
and $\mathfrak{R}^i(\boldsymbol{S}_0)$
for (\ref{es}), where $i=1,2,\dots$, originate from
the scaling symmetries for (\ref{sss}) and (\ref{es}),
\[
\ba{l}
\mathcal{S}_0=\left( \begin{array}{c}
3 t (p_{xxx}+9 p q p_x+3 p^2 q_x) +x p_x+p\\[2mm]
3 t (q_{xxx}+9 p q q_x+3 q^2 p_x) +x q_x+q\end{array}\right)\quad\mbox{and}\quad\\[10mm]
\boldsymbol{S}_0=\left( \begin{array}{c}
3 t \left(u_{xxx} - \ds\frac{3 u v_x u_{xx}}{u v + c} - \frac{9 u_x^2 v_x}{u v + c}
+ \frac{3 u^2 v_x^2 u_x}{(u v + c)^2}\right) +x u_x\\[5mm]
3 t \left(v_{xxx} - \ds\frac{3 v u_x v_{xx}}{u v + c} - \frac{9 v_x^2 u_x}{u v + c}
+ \frac{3 v^2 u_x^2 v_x}{(u v + c)^2}\right) +x v_x
\end{array}\right),
\ea
\]
respectively.

Finally, (\ref{es}) has  nonlocal symmetries
\[
\boldsymbol{G}_1=\left( \begin{array}{c}
\sqrt{u v+c}\exp(-w_1/2)\\[2mm]
0
\end{array}\right),\quad
\boldsymbol{G}_2=\left( \begin{array}{c}
0\\[2mm]
\sqrt{u v+c}\exp(w_1/2)
\end{array}\right),
\]
obtained by differentiating $\boldsymbol{Q}_{-1}$ with respect to $\omega_1$ and $\omega_2$,
but these symmetries are annihilated by $\mathfrak{R}$ and hence
do not lead to new hierarchies of nonlocal symmetries.

It would be interesting to find out whether the systems in question possess
nonlocal symmetries that do not belong to the above hierarchies,
which is the form of solutions invariant under the nonlocal symmetries
and whether these solutions could have any applications in nonlinear optics.

As a final remark, we note that because of the obvious symmetry of
the complex sine-Gordon~II system (\ref{csg2}) under the interchange
of $x$ and $y$, all of the above results concerning the recursion
operator, Hamiltonian and symplectic structures, and hierarchies of
symmetries for (\ref{csg2}) remain valid if we replace all
$x$-derivatives with $y$-derivatives and vice versa and swap the
operators $D_x$ and $D_y$. Interestingly enough, the recursion
operator $\tilde\mathfrak{R}$ obtained from $\mathfrak{R}$ upon such
an interchange proves to be inverse to $\mathfrak{R}$ on symmetries
of (\ref{csg2}) up to a constant factor. More precisely, for any
symmetry $\bi{K}$ of (\ref{csg2}) we have\looseness=-1
\[
\tilde \mathfrak{R} (\bi{K})=k^2 c^4\mathfrak{R}^{-1}(\bi{K}).
\]
Taking into account our earlier results we see that the `basic' hierarchy
of (\ref{csg2}) can be represented by a diagram of the form
\[
\cdots \rightarrow \mathfrak{R}^{-2}(\bi{u}_y)
\rightarrow \mathfrak{R}^{-1}(\bi{u}_y)
\rightarrow \bi{u}_y \rightarrow \bi{u}_x=\mathfrak{R}(\bi{u}_y)/kc^2
\rightarrow \mathfrak{R}(\bi{u}_x) \rightarrow \mathfrak{R}^{2}(\bi{u}_x)
\rightarrow \cdots,
\]
and we guess that {\em all\/} symmetries presented at this diagram are {\em local},
i.e., they do not involve nonlocal variables.
This is a fairly common situation for hyperbolic PDEs, see
e.g.\ the discussion in \cite{Bar}.
\looseness=-1

\section*{Acknowledgments}
The research of A.S. was partially supported by the Czech Grant Agency
(GA\v{C}R) under grant 
201/04/0538, and by the Ministry of Education,
Youth and Sports of the Czech Republic (M\v SMT \v CR) under grant MSM 4781305904.


\end{document}